# Manifestation of constrained dynamics in a low pressure spark

S K H Auluck

Physics Group, Bhabha Atomic Research Centre, Mumbai 400085, India

e-mail: skauluck@barc.gov.in; skhauluck@gmail.com

Abstract: Some features of neutron emission from dense plasma focus suggest that the participating deuterons have energy in the range of $10^5$ eV and have a directionality of toroidal motion. Theoretical models of these devices assume that the plasma evolves through a purely irrotational flow and thus fail to predict such solenoidal flow on the scale of the plasma dimensions. Predictions of a relaxation theory are consistent with experimental data [S K H Auluck, Physics of Plasmas,18, 032508 (2011)], but the assumptions upon which it is based are not compatible with known features of these devices. There is thus no satisfactory theoretical construct which provides the *necessity* for solenoidal flow in these devices. This paper proposes such theoretical construct, namely, the principle of constrained dynamics, and describes an experiment which provides support for this idea. The experiment consisted of low inductance, self-breaking spark discharge in helium at a pressure ~100 hPa between two pointed electrodes separated by 30-50 mm distance kept inside a vacuum chamber mounted on a low inductance high voltage capacitor. The current derivative signal showed reproducible sharp dips at *all* the extrema of the damped sinusoidal discharge. A planar diamagnetic loop centered with and perpendicular to the discharge axis consistently showed a signal representing rate of change of axial magnetic flux. The discharge plasma was very weakly ionized. Its acceleration was constrained by viscous drag of the neutrals, pressure gradient was constrained by heat conduction by neutrals and at the same time, the axial current density and azimuthal magnetic field were constrained to follow an oscillatory temporal profile. Under these conditions, radial momentum balance equation cannot be satisfied unless the plasma possesses a degree of freedom, which supplies the shortfall in momentum balance. Azimuthal symmetry of the plasma allows azimuthal current density to provide such degree of freedom. A qualitative explanation of observed phenomena is obtained using a simple model.

PACS: 52.50.Dg, 52.80.Mg; 52.80.Tn

Keywords: Plasma focus, accelerated ions, irrotational flow, toroidal motion, axial magnetic field, plasma relaxation

I. **Introduction**:

The role of accelerated ions in the neutron emission from dense plasma focus (DPF) and many other z-pinch devices is well known [1]. These accelerated ions have trajectories spanning the extent of the hot plasma inferred from soft x-ray images and possess a highly non-trivial directionality. The following studies provide essential data for understanding this directionality.

- Milanese and Pouzo [2] showed that typical neutron time-of-flight (ToF) spectra measured using a 128 m flight path at 90° to the axis on the Frascati 1-MJ plasma focus had 3 peaks: one central peak centered on the 2.45 MeV reaction Q-value of D-D fusion reaction with half-widths of 0.2-0.5 MeV and two well-separated lateral peaks. Such peaks were consistent with "a 100 keV deuteron loop in the plane determined by the gun axis and the observational direction" [2]. Similar vortex motion in the (r,z) plane has been revealed in numerical simulation models of Imshennik et. al. [3]. Sometimes, the lateral peaks were not well-resolved, resulting in a single broad peak with a width in the range 0.5-0.7 MeV [2]. Similar broad peaks were observed by Conrads et.al. [4] and Bernard et.al. [5] in dense plasma focus. Such broad side-on neutron spectra have also been observed in a frozen deuterium fiber z-pinch [6] and a gas-puff z-pinch [7].

- Neutron ToF spectra at 0° and 180° to the axis consistently show a shift towards higher and lower energy respectively [8, 9, 10, 11]. A similar observation is also made in a gas puff pinch [12].

- At the same time, clear evidence for existence of neutrons with energy higher than 2.45 MeV in the backward direction, showing presence of accelerated deuterons at 180° to the axis, is also available [13,14,15].

- Space-resolved neutron [16] spectra from laterally mirror symmetric points in the plasma show oppositely directed shifts. This has been interpreted as "the evidence of a 100 keV $d^+$ stream orbiting in a plane perpendicular to the experiment main axis" [16].

- Space resolved fusion proton spectra [17] from laterally mirror symmetric points in the plasma show a similar feature.

- Reconstruction of reaction proton images in the device coordinate system [17] shows un-ambiguous evidence for existence of azimuthal current comparable in magnitude with the axial current in the pinch zone at the time of neutron production.

- Existence of axial magnetic field in the plasma focus current sheath has been experimentally confirmed [16, 18].

Standard theories of DPF and other z-pinches [1, 19] always model the z-pinch in terms of a purely irrotational plasma flow, which fail to account for these features of the directionality of accelerated ions.

On the other hand, toroidal flow topology suggested by a Hall-MHD relaxation model [20] provides a unique descriptive fit [21] to all these observations. This toroidal topology is in fact embedded in the structure of azimuthally symmetric Chandrasekhar-Kendall (CK) function (curl eigenfunction in cylindrical geometry), which describes a purely solenoidal flow. The incompatibility between the traditional, purely-compressive irrotational flow description and the good descriptive agreement between experimental data and solenoidal flow model, leads to the question of applicability of relaxation theories in the context of the dense plasma focus and similar pinch devices.

Relaxation theories [20, 22, 23] are typically proposed to model self-organizing behavior of plasmas such as toroidal z-pinches, spheromaks and field-reversed configurations, which are bounded by close-in conducting walls. Current induced in these walls plays a crucial role in their dynamics. In contrast, dense plasma focus and many other linear z-pinch configurations do not have a close-in conducting wall and wall currents play no role. The relaxation theories are also organized around the idea that the plasma has (or should have) a tendency to evolve towards a minimum energy state subject to certain integral constraints [24]. Yoshida and Mahajan [25] have pointed out that there is some arbitrariness in choosing the energy minimization principle.

There is thus a conceptual vacuum in understanding the *necessity* of solenoidal flow in the dynamics of DPF and z-pinches. There is no satisfactory answer to the question: "Why should there be any solenoidal flow on the gross scale length of the plasma in purely compressive phenomena like DPF and z-pinches, which are carefully prepared to have good initial symmetry?" Turbulence, or short scale length solenoidal flow, may occur due to viscosity but that does not provide an answer.

In this paper, another theoretical construct is proposed and demonstrated using experimental data. This is based on the idea of "constrained dynamics", which is developed in the next section. Section three describes the experimental arrangement. Section four describes experimental observations. Section five discusses some question raised by the observations. Section six interprets the observations in terms of the idea of constrained dynamics. Section seven presents the summary and conclusions.

II. **Constrained dynamics**:

The focus of this section is on describing a principle which seeks to explain why gross solenoidal flow on the scale length of plasma dimensions must *necessarily* exist in certain plasma experiments, even if initial plasma preparation ensures complete absence of external solenoidal forces. This principle is, however, not *sufficient* to determine the detailed nature of the solenoidal flow.

The idea of constrained dynamics is based on the fact that terms of the momentum conservation equation are governed by widely dissimilar and independent processes. For example, pressure gradient is governed by processes which add or subtract energy (such as Ohmic heating, thermal conduction, radiation etc.) and particles (such as ionization and recombination) to the plasma and by irrotational flow which redistributes plasma particles in space. In contrast, a part of the magnetic (Lorentz) force is governed by current supplied by an external power source, which may impose a temporal profile on the magnetic force different from that of the pressure gradient. The resultant of these should ordinarily govern the plasma acceleration. But in dense plasma focus and many other kinds of z-pinches, the fully ionized current carrying plasma is *constrained* to move through an initially neutral medium. The viscous drag offered by the neutral medium to the magnetically accelerated plasma should result in a nearly constant drift velocity and a negligible acceleration. Streak camera pictures of the radial collapse phase of the dense plasma focus confirm this [26,27,28].

Under this condition, momentum balance can be satisfied only if there exists a degree of freedom in the dynamics, which can instantaneously supply the shortfall in the momentum balance equations produced as a result of different phenomena influencing different terms. This degree of freedom is in fact related to the absence external solenoidal forces, which ensures that the center of mass of the plasma can have only irrotational motion. This plays an essential role in the discussion.

When electron and ion fluid velocities $\vec{v}_e, \vec{v}_i$ are combined into a center-of-mass velocity $\vec{v} = (m_i \vec{v}_i + m_e \vec{v}_e)/(m_i + m_e)$ and current density, $J = en(\vec{v}_i - \vec{v}_e)$, the two-fluid equations of conservation of momentum for a fully ionized quasi-neutral plasma can be written down [29] as a center-of-mass conservation-of-momentum equation

$$\rho\left(\frac{\partial \vec{v}}{\partial t} + (\vec{v} \cdot \vec{\nabla})\vec{v}\right) = \vec{J} \times \vec{B} - \vec{\nabla}p - \vec{\nabla} \cdot \left(\frac{\vec{J}\vec{J}}{\varepsilon_0 \omega_p^2}\right) \qquad 1$$

and a generalized Ohm's Law.

The absence of external solenoidal forces ensured by initial plasma preparation process implies that the center-of-mass velocity is purely irrotational. Quasi-neutrality of the plasma ensures that the current density is divergence-free, i.e. solenoidal. It is reasonable to assume that the current would flow on surfaces of equal density (i.e. would be normal to density gradient). In a dense plasma focus, viscous drag offered by the neutral gas ahead of the sheath reduces the acceleration to zero. Putting all these observations together, 1 can be rewritten as

$$\rho \frac{1}{2}\vec{\nabla}|\vec{v}|^2 + \vec{\nabla}p + \frac{m_e}{ne^2}\frac{1}{2}\vec{\nabla}\left(|\vec{J}|^2\right) = \vec{J} \times \vec{B} + \frac{m_e}{ne^2}\vec{J} \times \vec{\nabla} \times \vec{J} \qquad 2$$

*This equation must necessarily be obeyed at all times and at all points.* Note that only a part of the current density, whose flux at the electrode surface equals the external current, is controlled by the external circuit; this may not be sufficient to balance equation 2. The remaining part of current density, whose flux at the electrode surface is zero, may exist in the form of localized current loops and is essentially a free parameter. Hence the terms proportional to electron mass in 2 cannot be neglected ab initio, since the "free" current density could, in principle, have a value inversely proportional to the electron mass. Null points of magnetic field may exist in case there are current loops; at these points, the electron inertia term on the RHS of 2 would play a crucial role in ensuring momentum balance.

These free currents are related to solenoidal flow of ions because of the irrotational character of the center-of-mass velocity. The solenoidal velocity of ions is related to solenoidal velocity of electrons by

$$\vec{v}_i^{sol} = -\frac{m_e}{m_i}\vec{v}_e^{sol} \approx \frac{m_e}{m_i}\frac{\vec{J}_{free}}{ne} \qquad 3$$

The constrained dynamics of the plasma focus produces a mismatch between the externally controlled magnetic force on the one hand and the gradient of thermal and Bernoulli pressure terms controlled by plasma evolution processes on the other hand. The greater the mismatch, the larger is the free current and the greater the solenoidal ion velocity. Constrained dynamics thus *necessarily* produces solenoidal ion motion on the gross scale length of the plasma.

An experimental demonstration of this principle should strive to create as large a mismatch between the externally controlled magnetic pressure and the plasma thermal and Bernoulli pressures. One way is to force the current to undergo oscillations and the pressure to be limited to very small values. This is described in the next section.

### III. **Experimental arrangement**:

The experiment consisted of a low inductance vacuum spark discharge arrangement adapted to provide a simple, microsecond duration helium plasma source for generating initial experience with plasma spectroscopy. The schematic of the experiment is shown in fig. 1. A vacuum chamber, 120 mm in both diameter and height, with 4 diagnostic ports was mounted to serve as a coaxial current return path with low inductance electrical connections on a 25 nH, 50 kV, 2.8 µF capacitor. Two pointed electrodes coaxially mounted within the chamber were provided for the spectroscopic source experiment: one was directly mounted on the high voltage bushing of the capacitor and the other was a movable electrode facing the high voltage electrode and was electrically connected to the vacuum chamber cum current return path. Adequate care was taken to prevent ground loops either through the charging circuit, or diagnostic connections or the vacuum system.

A magnetic probe consisting of a single 3 mm diameter loop of 0.5 mm thick insulated copper wire created at the end of RG174 cable and insulated with epoxy was installed at a fixed distance of 70 mm from the axis of the device for measurement of the current derivative (dI/dt) as a standard diagnostic. A diamagnetic loop centered with and perpendicular to the device axis of symmetry was also placed in the chamber to look for spontaneously generated axial magnetic field suggested earlier [30]. Initially, this consisted of a loop of stiff copper wire of nominal diameter 80 mm; later it was replaced with a copper laminated epoxy sheet machined so that its outer diameter had a snug fit to the inner wall of the vacuum chamber and its inner diameter was 76 mm. The copper lamination was machined out leaving a ring of 2 mm width, 80 mm inner diameter and 15 μm thickness. A 0.5 mm wide gap was created in this ring and a 50-Ohm coaxial cable was soldered across it with minimum lead length. The epoxy sheet was supported on three nylon rods of equal height kept on the insulating base plate of the chamber which maintained its plane exactly perpendicular to the device axis without any tilt. The diamagnetic loop was covered with a layer of epoxy resin to prevent currents due to photoemission from the metallic portion. The possibility of currents due to energetic charged particles emitted by the plasma and incident on the copper ring was ruled out by making observations with and without a 1 mm thick roll of mylar sheet inserted between the plasma and the diamagnetic loop.

The chamber was evacuated to $10^{-5}$ hPa with an oil diffusion pump and then filled with helium at various pressures between 1 and 125 hPa, which was the operating range of the Edwards capsule dial gauge used for pressure measurements. All the data reported in this paper, however, is for 100 hPa of helium. The capacitor was charged until a self breakdown occurred at a voltage between 2-5 kV depending on the electrode separation and the fill pressure. When the two electrodes were brought very close (~0.5 mm), the time period of the conventional damped sinusoidal discharge waveform was consistent with a circuit inductance of 60±3 nH. The calculated contribution to this inductance from the 20 mm diameter movable electrode extended about 30-50 mm beyond its normal position was 14±4 nH, The inductance of discharge circuit beyond the inter-electrode gap was thus 46±5 nH.

The discharge was visible to the unaided eye as a straight cylindrical luminous column about 10-15 mm in diameter and 30-50 mm long. Framing camera pictures using Imacon 790 camera at speeds of 5 x $10^6$ frames per second and 1 x $10^6$ frames per second showed that the luminous column maintained its straight cylindrical shape throughout the discharge [31]. Starting from a diameter of 0.4 mm, the column continuously expanded at the rate of 3 mm/μs [31]. Time-and-line-of-sight integrated spectroscopy in helium indicated that temperature, estimated from line intensity ratio method, was in the range of 0.5 eV at a filling pressure of 100 mbar. The electron density calculated from Saha equation for these numbers comes out to $1.7 \times 10^{12}$ electrons/cm$^3$ corresponding to a degree of ionization ~$6 \times 10^{-7}$ [31]. This number for density is a very crude order of magnitude estimate, since it was based on a time integrated measurement without any spatial resolution. There are reasons to believe (discussed in the next section) that the current carrying channel had a much higher density, although there is no direct evidence concerning its numerical magnitude. The spectroscopic estimates probably represent a global average.

## IV. Observations:

Peculiar current derivative signals were observed when self-breakdown between the two pointed electrodes was arranged.

Fig. 2 shows two examples of the dI/dt signal showing sharp dips at each extremum. These signals were found to be quite reproducible. Often, the dI/dt signals obtained on two different days under identical conditions of pressure were found to be *exactly* identical. The voltage across the capacitor, measured with a 40 kV, 75 MHz, Tektronix high voltage probe is shown in Fig. 3 along with the dI/dt signal *from the same shot*. The voltage signal in Fig 3 was recorded on an analog storage oscilloscope and was limited by its writing speed as evidenced by the dark patches seen on the trace in its initial portion. The absence of similar dark patches at the extrema of *this signal* indicates that dips of the type seen in the corresponding dI/dt signal are not present. Even when recorded on a digital storage oscilloscope with better writing speed, the voltage signal never showed any dips at the extrema of the damped sinusoidal waveform. In very rare cases, a small, sharp *spike* could be observed riding on a single extremum of the voltage waveform. The 3.5±0.1 µs time period of the damped sinusoidal voltage waveform was consistent with a circuit inductance of 105 ± 10 nH. In the operating voltage range (2-5 kV), the peak current is estimated to be in the range 8-22 kA.

If this entire current were to be carried by a plasma with the size and density mentioned above, it would imply electron drift velocity comparable to the velocity of light. There is no basis to expect such high velocities at such small values of voltage and current. It would be more likely that the current would be carried by a spatially more localized streamer channel embedded within a less-ionized plasma, having a much higher linear density than indicated by the spectroscopic estimates. There is no direct evidence concerning this. The only significance that may be attached to the spectroscopic estimate is that it probably represents a global average.

In a continuous series of shots without changing the fill gas, the dI/dt signal was found to change gradually. But in all cases, its chief characteristic, namely, the discontinuities at the extrema, was observed without fail. When many shots were taken in rapid succession (nearly one per 5 seconds), about 30% of the shots showed a great deal of distortion in this typical dI/dt signal. By decreasing the sensitivity of the measurement from 2 V/div to 200 V/div using attenuators, it was found that giant sharp transients occurred in these cases with peaks one to two orders of magnitude larger than the typical "clean" signal. (fig. 4). In subsequent discussion, the noisy shots in a rapid sequence of shots are referred as N type shots and other shots are referred as C (for "clean") type shots.

In each C type shot, a signal (referred as $\dot{\Phi}_z$ signal) was observed in the diamagnetic loop in every shot. This signal was, however, not reproducible to the same extent as the dI/dt signal. Fig. 5 shows two examples of the $\dot{\Phi}_z$ signals together with the corresponding dI/dt signals.

In N type "noisy" shots, the $\dot{\Phi}_z$ signals were also very large. Very interestingly, in the "noisy" shots, the $\dot{\Phi}_z$ signal had a long lived component sometimes lasting for 2 milliseconds while the dI/dt signal never lasted for more than 10 μs (Fig. 6). By using three oscilloscopes, first one recording the dI/dt signal at 200 V/div, 1 μs/div, second recording $\dot{\Phi}_z$ at 20 μs/div and the third recording the same $\dot{\Phi}_z$ signal (using a matched T-splitter) at 500 μs/div, it was established that the long lived $\dot{\Phi}_z$ signal was occurring only when the sharp transients were present in the dI/dt signal. In one experiment, the $\dot{\Phi}_z$ recording oscilloscope was triggered by the dI/dt signal and it was found that the large $\dot{\Phi}_z$ signal began after the transients of dI/dt signal ceased (fig. 7).

## V. Discussion:

These observations raise a number of questions which are discussed below

1. Are the observed signals artifacts of instrumental construction or do they indicate real phenomena?

Pulse power professionals dealing with large currents and voltages often observe similar signals in their b-dot probes and attribute the "distortion" to dielectric breakdown in the probe. In the present case, the voltage across the probe is directly fed to the oscilloscope and is measured to be of the order of 1 volt or less, not sufficient to cause dielectric breakdown in the epoxy-insulated probe. One could suspect that the dips at the extrema of the dI/dt signals could be caused by interface phenomena (such as semiconducting oxide layers acting as nonlinear circuit elements) at the junction of the plasma and the electrodes or at the various electrical joints in the discharge circuit. However, it was observed that these dips gradually diminished and vanished altogether when the pressure in the chamber was increased to 600-1000 hPa, as read on the gauge attached to the helium cylinder. This observation rules out connections outside the chamber as being responsible for the dips. Initial observations were made with helium; but later it was found that similar signals were obtained with air. This argues against any plasma/electrode interaction of chemical origin as a cause of these phenomena. Langmuir sheath phenomena at the plasma-electrode interface cannot cause $\dot{\Phi}_z$ signals. These considerations argue in favor of the observations being due to real plasma phenomena.

2. Are the $\dot{\Phi}_z$ signals artifacts of improper measurement or real?

The diamagnetic loop was extremely planar, centered with and perpendicular to the axis. The signal that was measured at the cable was the integration of the electric field along the copper track, whose two ends were *at the same axial location*. Only azimuthal component of electric field could therefore contribute to the signal. In principle, the joint between the cable and the copper ring could introduce some contamination from the axial component of electric field. But the cable was soldered with minimum lead length in order to minimize likelihood of such contamination. The fact that the $\dot{\Phi}_z$ and dI/dt signals had a strikingly different temporal structure in at least some shots (see fig. 5 (b)) shows that there was no or negligible cross-sensitivity between the two. The signal was not affected by inserting a 1 mm thick roll of mylar sheet between the plasma column and the diamagnetic loop showing that there was no effect due to energetic particles from the plasma creating a signal in the diamagnetic loop. Similarly any electrical signal from photoemission from the metallic surface of the diamagnetic loop was ruled out by covering the copper ring and the solder joint between the ring and the cable with an epoxy coating.

3. Can known plasma phenomena create a $\dot{\Phi}_z$ signal in this experiment?

In principle, a current-carrying plasma could have kink or flute oscillations having a helical structure, which could be accompanied by axial magnetic flux. Such oscillations would have a natural frequency related to Alfven transit time, which would be too fast to be measured by the oscilloscope with 100 MHz bandwidth. Perhaps the envelope of such modes, related to the natural frequency of the LCR discharge circuit powering the plasma, could be within the detection bandwidth. The significant aspect of these observations is the occurrence of discontinuities at the extrema of the current derivative signal. In at least one shot (see fig 5 (b), the $\dot{\Phi}_z$ signal suddenly decreased all the way to the baseline and again recovered in the same polarity. Such sudden excursions occurring on a faster time scale than that of the discharge indicate that some kind of nonlinear phenomenon is coming into play. Instability of helical kink modes may explain one isolated singular event. But the observations indicate a *periodic train* of such singular events occurring reproducibly in many shots. Unstable growth of kink or any other mode therefore seems to be ruled out as the cause of the observed periodically repeated, short timescale features of $\dot{\Phi}_z$ signals.

The long lasting $\dot{\Phi}_z$ signals in N-type shots suggest formation of a self-sustaining plasma-magnetic field configuration. There is insufficient data to discuss relevance of any particular plasma phenomenon to this observation.

4. What does the difference between the nature of voltage and current derivative signals indicate?

   The fact that voltage measured across the capacitor, which should equal the sum of the inductive and resistive voltage drops, does not show the dips at the extrema and the dI/dt signal in the same shot does show them, indicates presence of a time-varying inductance L. The current I is close to zero at these times and therefore the Ohmic voltage drop cannot have any influence on the voltage signal at these times. The non-appearance of dips in the voltage signal therefore indicates that the derivative of magnetic flux LI has no discontinuity. If dI/dt has discontinuities and d(LI)/dt does not, this implies that the inductance L must have a time-dependent structure related in some manner with the current.

5. Is it possible to identify some critical condition which can be repeatedly met and "unmet" during the damped sinusoidal excursion of the current?

   For the order of plasma density inferred from spectroscopy, the Bennett current $I_B$ [32] required to obtain equilibrium between plasma pressure and magnetic pressure at a plasma radius of 1 cm is less than 1 ampere. In contrast, the discharge current is of the order of kilo-amperes. The "pressure equilibrium condition" would occur when the discharge current I(t) equals $I_B$, which would occur near the zeros of I(t), which means, at the extrema of dI/dt.

## VI. Interpretation in terms of constrained dynamics:

This section discusses how the observed phenomena can be interpreted in terms of the principle of constrained dynamics.

The plasma is known to be very weakly ionized. Therefore, the motion of the ion component is expected to be mainly governed by collisions with the neutrals. The motion of neutrals consists of the observed hydrodynamic expansion at the rate of 3 mm /μs. This can be taken as adiabatic variation of plasma radius on the time scale of the dI/dt dips, *with negligible radial acceleration*. From the global estimate of average electron density inferred from spectroscopy, the estimated collision-less electron skin depth $c/\omega_{pe} \sim 4$ mm, *of the same order as the radius of the luminous column observed in framing pictures*. The electron momentum convection term in the equation of momentum conservation 1 is then non-negligible. Using the assumption of azimuthal symmetry $(\partial/\partial\theta = 0)$, justified by framing camera evidence and neglecting end effects $(\partial/\partial z \approx 0)$, the radial component of equation 1 can be written as

$$J_\theta B_z - J_z B_\theta - \frac{\partial p}{\partial r} + \frac{J_\theta^2}{r\varepsilon_0 \omega_{pe}^2} = \rho\left(\frac{\partial \vec{v}}{\partial t} + (\vec{v} \cdot \vec{\nabla})\vec{v}\right)_r \approx 0 \qquad 4$$

This equation provides the backdrop for appreciating the peculiar circumstances of the reported experiment.

- The axial current density $J_z$ clearly depends on the current supplied by the capacitor discharge circuit and on the plasma radius determined by the hydrodynamic motion of neutrals. It is *forced* to undergo damped sinusoidal oscillations at the natural frequency of the LCR circuit. The azimuthal magnetic field is tied with this behavior of $J_z$ by Maxwell's equations.
- The pressure gradient is governed by resistive heat deposition by the current, the heat taken away by the neutrals, density and radius both governed by the hydrodynamic motion of the neutrals. *It cannot have an oscillatory form following that of the $J_z B_\theta$ term in 4.*
- The left hand side of equation 1 is the radial acceleration of the plasma. The motion of the ionized species driven by magnetic pressure into the viscous neutral gas may result in a constant terminal drift velocity. Any tendency of the ionized species to undergo oscillatory motion driven by the $J_z B_\theta$ magnetic force is expected to be damped by the collisions with neutrals.
- The only way equation 4 can be balanced is for the terms $J_\theta B_z + J_\theta^2 / r\varepsilon_0 \omega_{pe}^2$ to provide the shortfall arising from an oscillatory $J_z B_\theta$ term, a non-oscillatory pressure gradient term and a nearly-zero non-oscillatory radial acceleration term. *However, there is no phenomenon which supplies an initial value for the azimuthal current density $J_\theta$. This current must therefore be spontaneously generated.* Any inquiry into the detailed mechanism for this spontaneous generation would have to involve a discussion of cooperative interaction between random fluctuations [32] as in the case of ferromagnetism.

The azimuthal current density required to balance the momentum conservation equation 4 is given by

$$J_\theta = \frac{(r\omega_{pe}^2)}{2\mu_0 c^2} \left\{ \pm \sqrt{B_z^2 + \frac{4\mu_0 c^2}{r\varepsilon_0 \omega_{pe}^2} \left\{ J_z B_\theta + \frac{\partial p}{\partial r} + \rho g_r \right\}} - B_z \right\} \quad 5$$

$$g_r \equiv \left( \frac{\partial \vec{v}}{\partial t} + (\vec{v} \cdot \vec{\nabla}) \vec{v} \right)_r$$

It is reasonable to assume that the center of mass motion of the weakly ionized plasma is dominated by the dynamics of the neutral component, so that the radial acceleration is related to the radial gradient of the pressure $p_n$ of the neutral species:

$$\rho g_r = -\frac{\partial p_n}{\partial r} \quad 6$$

The Bennett profile [33] with a radial scale length $b^{-1}$ can be used for the density as a model example:

$$n = \frac{n_0}{\left(1+b^2r^2\right)^2} \qquad 7$$

Assuming a uniform axial drift velocity for electrons and uniform temperatures for all plasma species, the following relations are obtained, with $p_0$ representing the total pressure at r=0:

$$p + p_n = \frac{p_0}{\left(1+b^2r^2\right)^2}; J_z = \frac{Ib^2}{\pi}\frac{1}{\left(1+b^2r^2\right)^2}$$

$$J_z B_\theta + \frac{\partial p}{\partial r} + \rho g_r = \frac{\mu_0 b^4 r}{2\pi^2 \left(1+b^2r^2\right)^3}\left\{I^2 - \frac{8\pi^2 p_0}{\mu_0 b^2}\right\} \qquad 8$$

Maxwell's equations then give for $B_z$

$$-\frac{\partial B_z}{\partial r} = \frac{\left(r\omega_{pe}^2\right)}{2c^2}\left\{\pm\sqrt{B_z^2 + \frac{B_0^2}{\left(1+b^2r^2\right)^3}} - B_z\right\}$$

$$B_0^2 = \frac{2\mu_0^2 c^2 b^4}{\pi^2 \omega_{pe}^2}\left(I^2 - I_B^2\right); I_B^2 = \frac{8\pi^2 p_0}{\mu_0 b^2} \qquad 9$$

This equation is not invariant under the mirror transformation $\hat{z} \rightarrow -\hat{z}$ unless the sign of the radical is replaced by the projection of a physical unit vector on the $\hat{z}$ axis. Since axial current is the only physical vector available, the sign of the radical is chosen to be the sign of the current flow.

Equation 9 must be supplemented with the condition that the flux of $B_z$ over the radius of the metallic vacuum chamber, assumed to be a perfect conductor over the time scale of the experiment, should be constant and equal to zero. *This implies that $B_z$ must reverse its sign within the chamber.*

This condition can be met by a solution of equation 9 only when $I^2 - I_B^2 \geq 0$. For $I^2 - I_B^2 < 0$, the square-root would give imaginary values at the point where $B_z$ reverses its sign. Hence, as the current oscillates and becomes less than $I_B$, the axial magnetic field $B_z$ should become zero. The sharp drop of $\dot{\Phi}_z$ signal in Fig 5(b) is consistent with this.

Inductance of the discharge circuit can be defined in terms of the total magnetic energy. The energy associated with $B_z$ should therefore contribute to the inductance. Clearly, the magnitude of $B_z$ is related to $B_0$, which is proportional to $\sqrt{I^2 - I_B^2}$. The flux linked with the capacitor discharge circuit may then be postulated to have the form

$$\Phi = L_0 I + L_1 I \sqrt{1 - I_B^2/I^2} \qquad 10$$

The azimuthal current density must also be proportional to $\sqrt{I^2 - I_B^2}$ and should contribute to resistive dissipation. The model circuit equation then takes the form,

$$L_{eff} \frac{dI}{dt} = V_0 - C^{-1} \int_0^t I dt - R_{eff} I$$

$$L_{eff} \equiv \left( L_0 + L_1 \sqrt{1 - I_B^2/I^2} + \frac{L_1 I_B^2}{\sqrt{I^2 - I_B^2}} \right) \text{ for } I^2 - I_B^2 > 0, \, L_0 \text{ otherwise} \qquad 11$$

$$R_{eff} = R_0 + R_1 \sqrt{1 - I_B^2/I^2} \text{ for } I^2 - I_B^2 > 0, \, R_0 \text{ otherwise}$$

In 11, $L_0, L_1, R_0, R_1$ are adjustable model parameters. The rapid transients caused by the square root terms in 11 would be smoothened by the finite bandwidth of the recording system, which can be modeled by a probe which partly integrates the detected signal with a certain integration time constant. Fig. 8 shows such simulated probe signal, where the adjustable parameters have been selected to provide gross visual resemblance with the observed signals.

The simulation model reproduces the *occurrence* of dips in the dI/dt signal *but not their form*. In the observed signal, the amplitude does not regain the damped-sinusoidal trend line after the dip, while the simulated signal does. The simulation model is thus clearly not an adequate representation of reality.

The above discussion provides a plausible explanation for the observations reported above, particularly, the *existence* of dips at extrema in the current derivative signal, the existence of an axial magnetic field signal and absence of dips in the voltage signal. The significant point is the necessity of invoking the electron momentum convection term for dynamically balancing the momentum conservation equation, whose terms are governed by widely different phenomena, some of which have an oscillatory character while others are non-oscillatory. When the magnetic force due to the externally supplied current passes through zero in the course of oscillation, it is this term which ensures conservation of momentum. Azimuthal symmetry plays an important role in keeping the azimuthal electron drift current free from any driving force and thus making it a degree of freedom, which can bridge the deficit in momentum balance caused by the fact that every other term in the momentum balance is controlled by a different physical phenomenon. The fact that no plasma phenomenon provides a significant initial value for the azimuthal current density suggests the existence of a phase-transition-like behavior, where random fluctuations interact cooperatively [32] to generate a non-random observable effect analogous to the order parameter in ferromagnetic phase transition.

The principle of constrained dynamics would seem to suggest that azimuthal currents should *always* exist in a dense plasma focus. Indeed, a signal (See fig 9) was seen in a diamagnetic loop placed outside the cathode of a dense plasma focus and 10 mm above the anode in every shot in a series of few tens of shots [34]. It began with the start of current and continued well past the pinch phase.

This is a very different observation than the experimental observation of axial component of magnetic field in the plasma focus current sheath by Krauz et. al. [18] just before the pinch phase. The measurement of time derivative of axial magnetic field by Krauz et al is a differential local measurement; its interpretation and importance is related with observation of localized toroidal and helical structures by interferometry on the same device [35] and recent revelation that the neutron emission coincides with the increase and decrease of axial magnetic field [36].

In contrast, the detection of axial flux outside the current carrying region is more like an emission from the plasma as a whole: a global property, which reveals a new aspect of global dynamics of the dense plasma focus, which has not been suspected before. This concerns questions about the nature of boundary conditions [37] on the axial magnetic field. The standard boundary condition is that the azimuthal electric field must be zero on the continuous cylindrical metallic boundary, leading to conservation of axial flux within the vacuum chamber. This boundary condition, along with any theoretical model of azimuthal current density, would determine the absolute value of axial magnetic field at the axis *before* the pinch phase. One could deliberately introduce another azimuthally continuous cylindrical metallic dummy wall between the cathode and the vacuum chamber, which should change the boundary condition and hence the value of $B_z$ at the axis. This can be directly measured by introducing an optical fiber based Faraday rotation probe similar to that described in [38]. If the dummy wall changes the value of $B_z$ at the axis, it would imply that the plasma focus phenomenon is not dependent on the absolute value of spontaneous axial magnetic field but only on its derivative: a very surprising conclusion. But, if the value of $B_z$ at the axis is not changed by the introduction of the dummy wall, then what governs the boundary conditions?

Observations of Barrett [39] concerning topological effects in electromagnetism may need to be examined in this context. During the run-down phase, the plasma focus electrode geometry does not represent a simply-connected domain. Electromagnetism may have a different character in such geometry. For example, an arbitrary vector field may consist of a non-zero harmonic component in addition to an irrotational and a solenoidal component guaranteed by Helmholtz's Theorem.

The polarity of axial flux within the cathode diameter with respect to the polarity of current is another important aspect. It could be a fixed positive or negative polarity, a randomly varying polarity or a polarity which can be manipulated by operating parameters of the device such as pressure or nature of gas etc. The data of Jaeger and Herold [17 ] indicates a fixed polarity for the azimuthal current. The recent result [40] that magnetic moment of a stream of free electrons is aligned with its momentum vector may play an important role in determining this polarity and may act as the symmetry-breaking mechanism in a ferromagnet-like phase transition behavior.

## VII. Summary and conclusions:

This paper introduces the idea of constrained dynamics as a consequence of the fact that various terms in the momentum balance equations are governed by widely dissimilar and independent phenomena. Conservation of momentum then requires existence of a physical free parameter, which must attain a value sufficient to bridge the shortfall in momentum balance *at every instant at every point*. Electron momentum convection by solenoidal electron currents serves as a free parameter of this type. In the absence of external solenoidal forces, ensured by plasma preparation process, these solenoidal electron currents are necessarily accompanied by solenoidal ion flow. The principle of constrained dynamics therefore provides the conceptual *necessity* of existence of solenoidal ion motion in a purely compressive irrotational flow such as a dense plasma focus.

Observations of anomalous current flow in a low pressure spark experiment are described, which can be interpreted in terms of the idea of constrained dynamics. The experiment is in the nature of a gaseous discharge in low pressure (~100 hPa) gas acting as a self-breaking high voltage switch in a low inductance, high voltage capacitor discharge circuit. The anomaly consists of reproducible dips at the extrema of the current derivative signal measured by a b-dot probe, observation of axial magnetic flux derivative signals different in profile and behavior from the current derivative signals and voltage signals having a conventional damped-sinusoidal form without any distortion. There are other peculiar phenomena, such as large spikes in the current derivative signal and long-lived structure in the axial magnetic flux derivative signal, which occur with a lesser frequency.

Discussion of these observations suggests that they represent real plasma phenomena, where electron momentum convection term in the momentum balance equations is non-negligible and plays a part in spontaneous generation of azimuthal current and axial magnetic field. Further exploration of these phenomena is therefore likely to offer valuable insights in plasma dynamics.

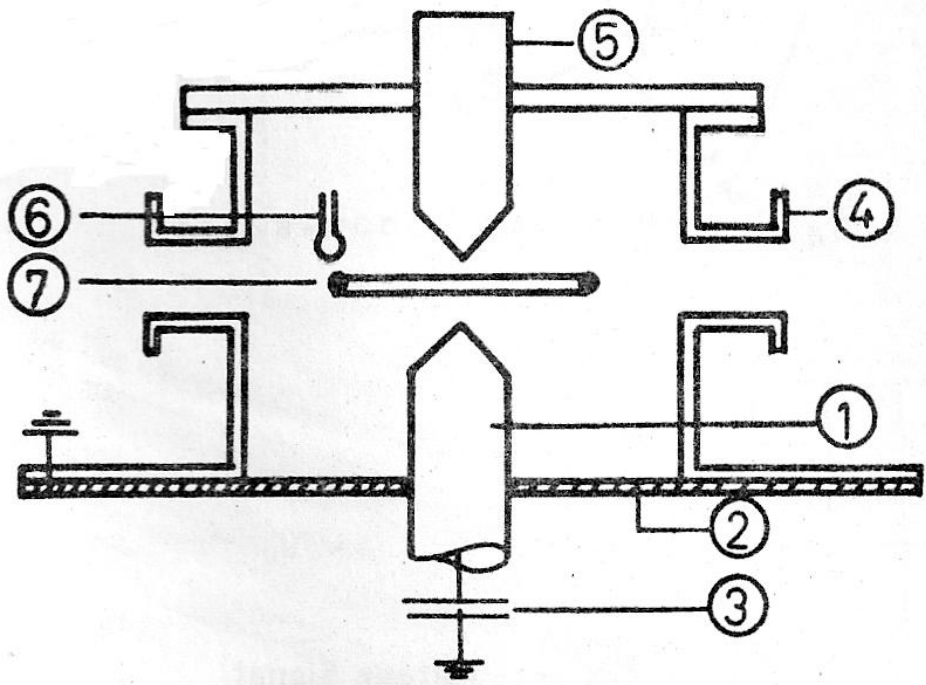

Fig. 1: Schematic of the experiment: (1) High Voltage Electrode (2) Insulator (3)Capacitor (4) Vacuum Chamber (5) Ground electrode (6) $\dot{B}_\theta$ probe (7) Diamagnetic loop for axial flux.

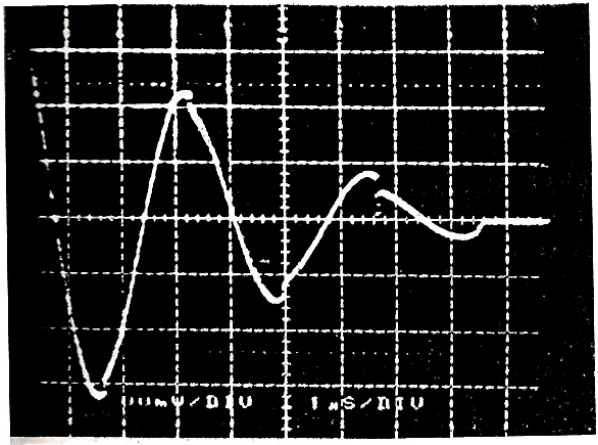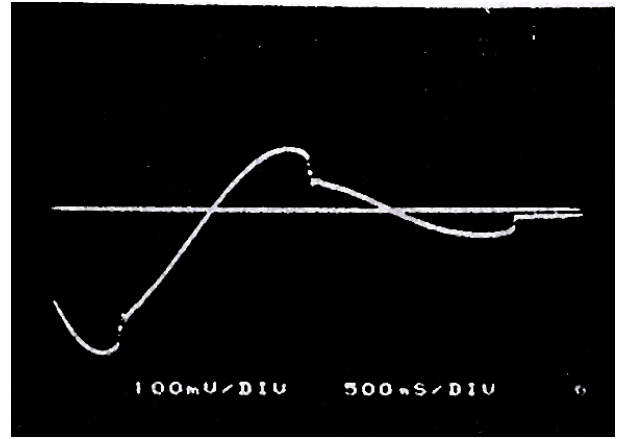

Fig. 2: Two examples of $\dot{B}_\theta$ signals

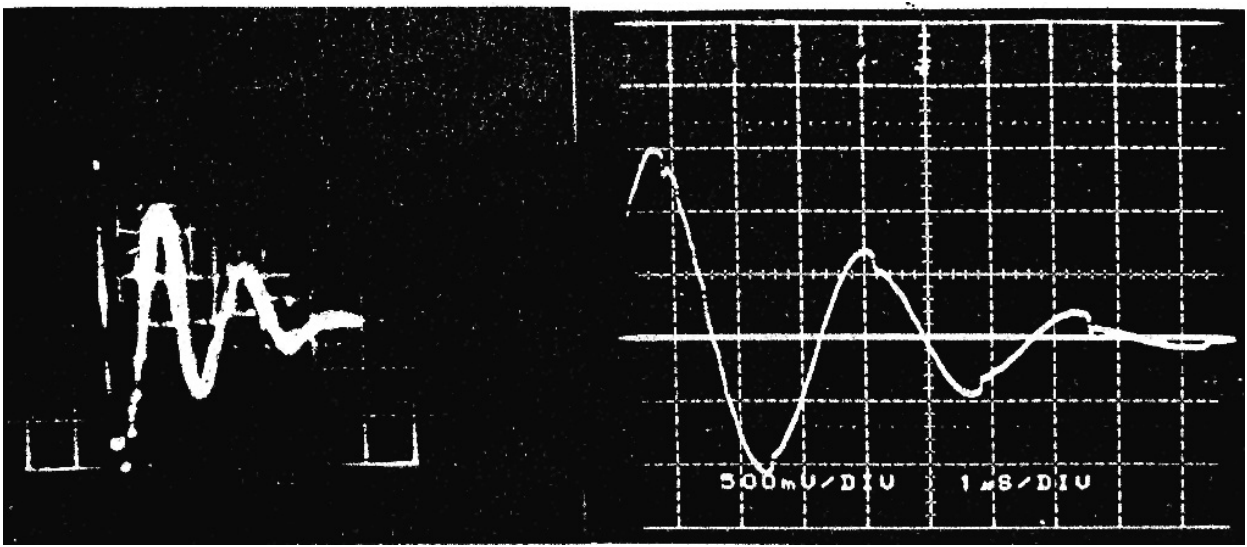

Fig. 3: Voltage signal measured by a Tektronix 40 kV 75 MHz probe along with $\dot{B}_\theta$ signal from the same shot.

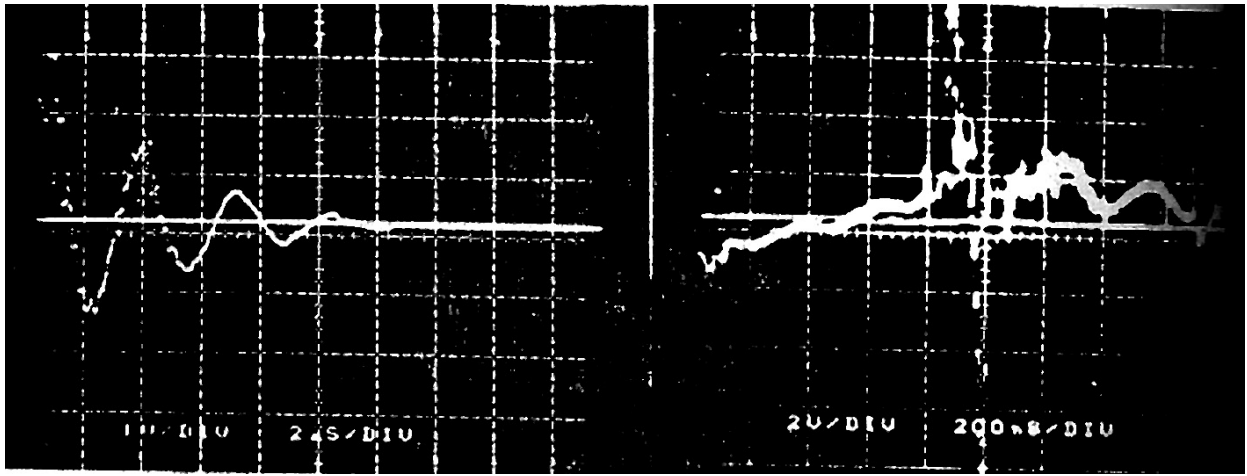

Fig 4. $\dot{B}_\theta$ signals from a noisy shot. The first signal is at a sensitivity of 100 V/div, 2 μs/div and the second is at a sensitivity of 200V/div, 200 ns/div using attenuators. The second signal illustrates the sharply varying transients.

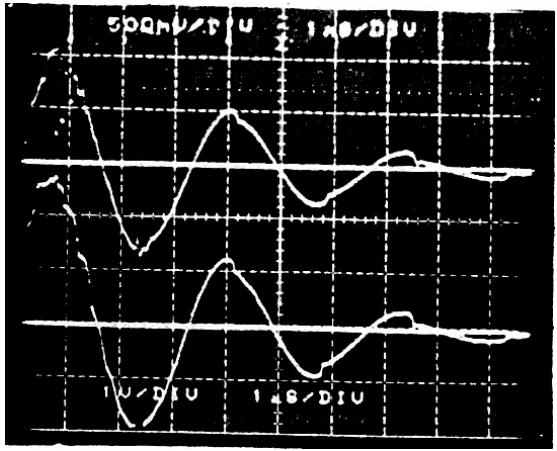
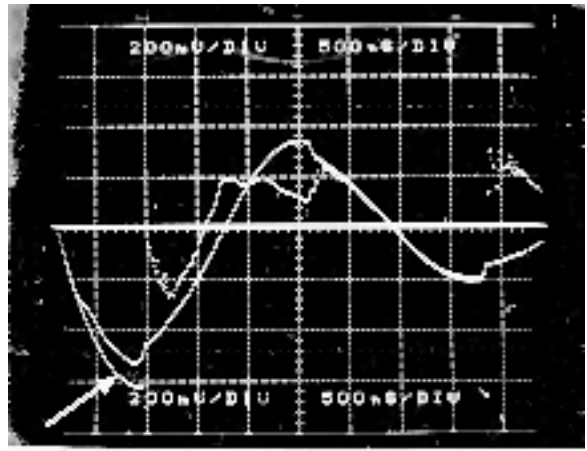

(a) (b)

Fig. 5: Two examples of $\dot{\Phi}_z$ signals. In (a), the top signal is $\dot{\Phi}_z$ and the bottom signal is $\dot{B}_\theta$. In (b), the $\dot{\Phi}_z$ signal is marked with an arrow. The striking difference in the temporal shapes of the two signals in (b) shows that there is no cross-contamination in the measurements of axial and azimuthal field components.

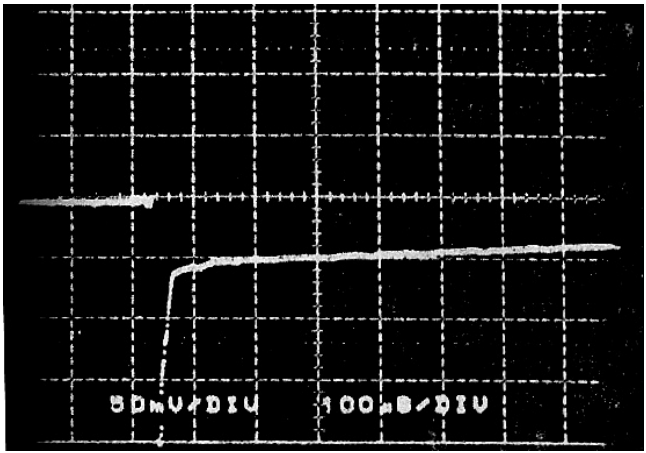
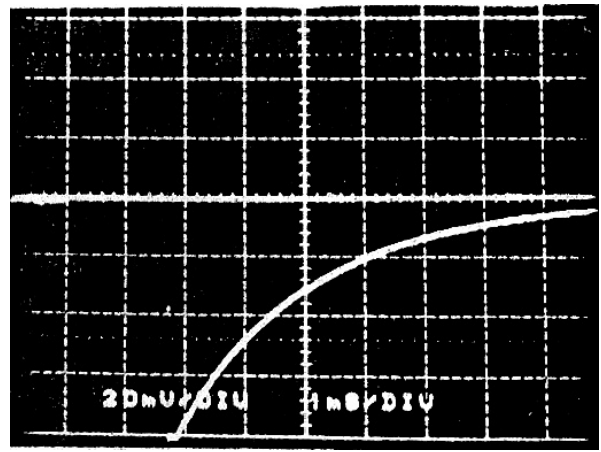

(a) (b)

Fig. 6. Long lived structure in a noisy shot: $\dot{\Phi}_z$ signal at 50 mV/div, 100 μs/div and the same signal seen at 20 mV/div, 1 ms/div on a different oscilloscope. The exponential decay time is 2.2 ms/ The signal actually goes up and comes down in a few microseconds (too rapidly to be recorded at this setting ) and then decays slowly.

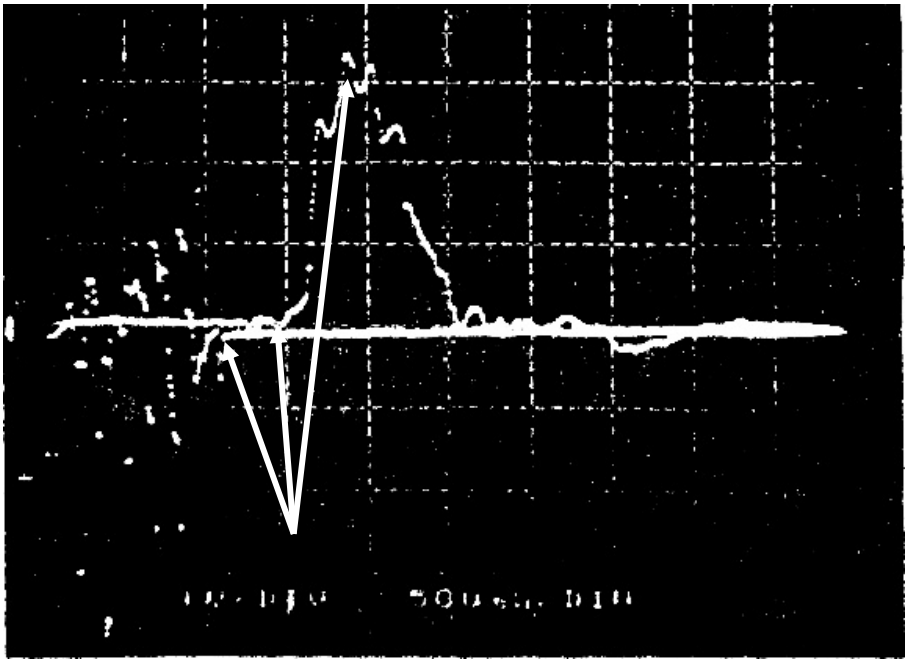

Fig 7. $\dot{B}_\theta$ signal at 200 V/div, 500 ns/div, and $\dot{\Phi}_z$ signal (indicated by arrows) at 100 V/div, 500 ns/div recorded *in the same noisy shot* on two different oscilloscopes but photographed on the same film. The oscilloscope recording the $\dot{\Phi}_z$ signal was triggered by the $\dot{B}_\theta$ signal. The noisy $\dot{B}_\theta$ signal reaches its baseline at about 1.1 μs, (shown by the first arrow). The $\dot{\Phi}_z$ signal rises from its baseline (shown by second arrow) well after that, at about 1.5 μs, reaches a peak (third arrow) and then goes to zero. Since this is rate of change of axial flux, the axial flux is seen to persist with a nearly constant amplitude for the duration of this trace (about 5 μs).

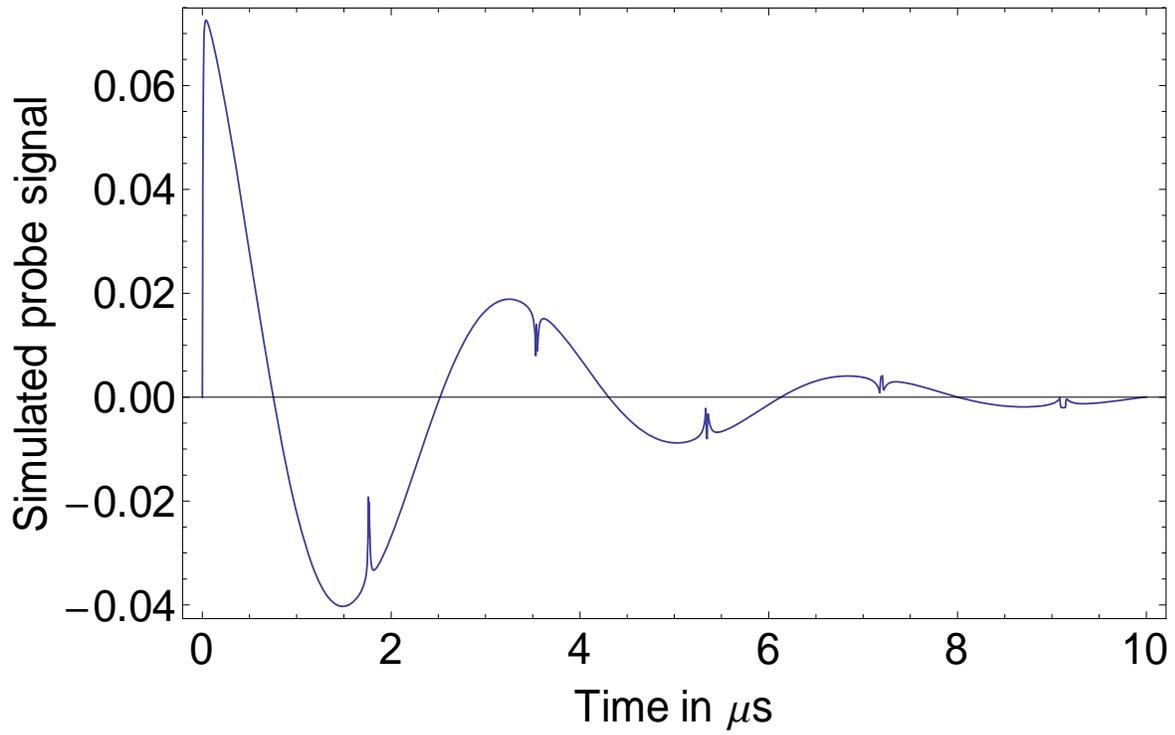

Fig. 8. Simulated probe signal. The following parameters are used: $L_0$=100 nH, $L_1$=4 nH, $I_B$=30 A, $V_0$=4 kV, $C_0$=2.8 µF, $R_0$=50 mΩ, $R_1$=40 mΩ. The probe integration constant is arbitrarily assumed to be 4 ns, reflecting bandwidth limitation of the entire set-up.

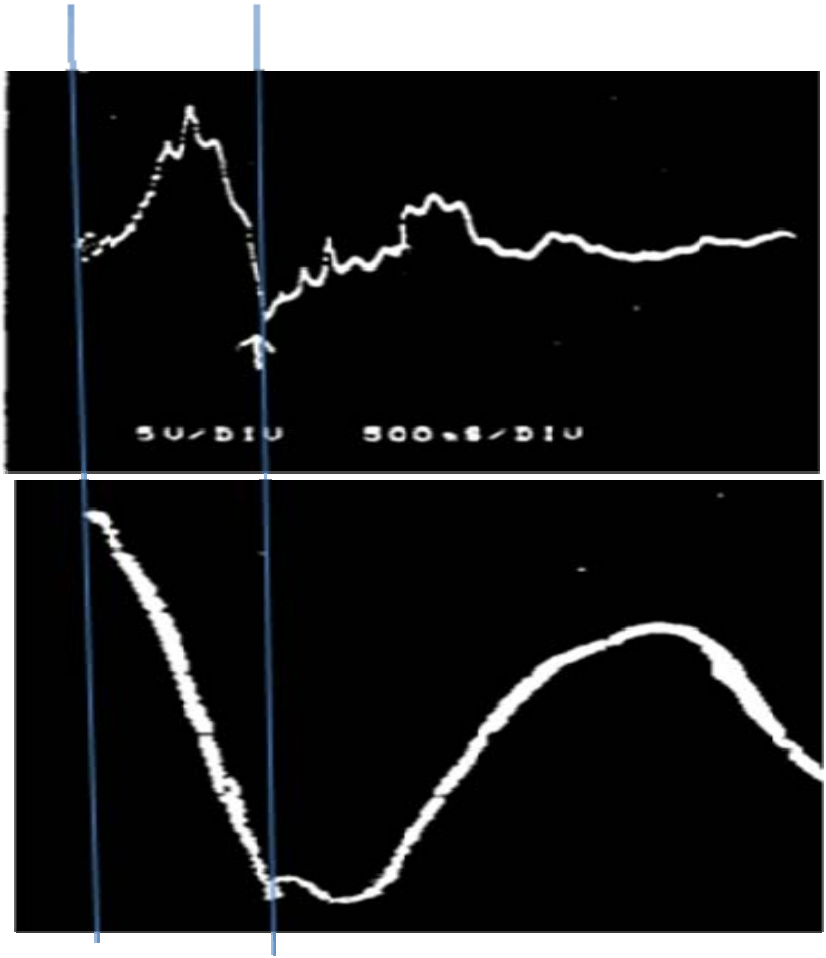

Fig. 9: Diamagnetic loop signal in a plasma focus (upper signal). Lower signal is the dI/dt signal from a Rogowski coil. Note that the diamagnetic signal begins at the same time as the discharge current. The plasma focus had a 22 mm diameter, 110 mm long anode, 86 mm outer diameter cathode consisting of 12 rods of 10 mm diameter and was operated at 70 kA peak current. The ring was placed at a height 10 mm above the anode and *was well outside the cathode*

Author Bio-sketch:

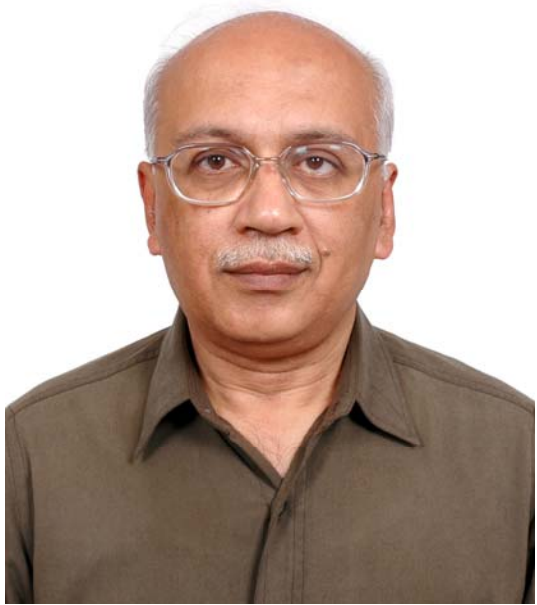

S.K.H. Auluck joined the Neutron Physics Division of Bhabha Atomic Research Center (BARC) in 1978, after graduating from the Indian Institute of Technology, Kharagpur and the BARC Training School. He worked as a Guest Scientist at the POSEIDON plasma focus facility at the Institute for Plasma Research at the University of Stuttgart from 1988 to 1990. He was awarded the Ph.D. degree by Bombay University in 1993 for his thesis "Spontaneous magnetic fields and related finite electron mass effects in fusion plasmas".  His interests lie in physics of the plasma focus and z-pinches, power concentration phenomena in various systems and some topics in optics. He has worked on the technological aspects of manufacturing special purpose capacitors, neutron sources and certain energetic devices. He is a Professor at the Homi Bhabha National Institute, Mumbai (a Deemed University) and Outstanding Scientist at the Physics Group, Bhabha Atomic Research Center, Mumbai. He has served as faculty at two Schools on Dense Magnetized Plasmas at ICTP, Trieste in 2010 and 2012. He is the Representative of India on the International Scientific Committee for Dense Magnetic Plasmas attached with the International Center for Dense Magnetized Plasmas, Warsaw, Poland under an agreement between UNESCO and the Polish National Atomic Energy Agency.